\RequirePackage[2020-02-02]{latexrelease}
\documentclass[twocolumn]{revtex4-1}
\usepackage{graphicx}
\usepackage{graphicx,booktabs,array}
\usepackage[latin1]{inputenc}
\usepackage{epsfig}
\usepackage{amsmath}
\usepackage{amsfonts}
\usepackage{float}
\usepackage{amssymb}
\usepackage[T1]{fontenc}
\usepackage{amssymb}
\usepackage{feynmp}
\usepackage[colorlinks=true
,urlcolor=blue
,citecolor=blue
,linkcolor=blue
,pagecolor=blue
,linktocpage=true
,pdfproducer=medialab
]{hyperref}
\usepackage{subfigure}
\usepackage{longtable}
\usepackage{multirow}
\usepackage{notoccite}
\usepackage[nodisplayskipstretch]{setspace}
\usepackage{placeins}
\graphicspath{{./Diagrams/}}
\def\be{\begin{equation}}
\def\ee{\end{equation}}
\def\bea{\begin{eqnarray}}
\def\eea{\end{eqnarray}}

\def\lsim{\raise0.3ex\hbox{$\;<$\kern-0.75em\raise-1.1ex\hbox{$\sim\;$}}}
\def\gsim{\raise0.3ex\hbox{$\;>$\kern-0.75em\raise-1.1ex\hbox{$\sim\;$}}}

\usepackage{pstricks}
\usepackage{epsfig}
\usepackage{xcolor}
\begin{document}
	
	\title{Explaining the ${\mathcal{R}}(D)$ and ${\mathcal{R}}(D^{\ast})$
		Anomalies \\ 
                     in the $B-L$ Supersymmetric Standard Model with Inverse Seesaw}
	\author{Dris Boubaa$^{1,2}$, Shaaban Khalil$^{3}$, Stefano Moretti$^{4,5}$ and Cem Salih Un$^{6,7}$} 
	\affiliation{$^{1}$Laboratoire de Physique des Particules et Physique Statistique, Ecole Normale Sup\'erieure-Kouba, 
		B.P. 92, 16050, Vieux-Kouba, Algiers, Algeria}
	\affiliation{$^{2}$Department of Matter sciences, Faculty of Science and Technology,
		Abbes Laghrour University Of Khenchela, BP 1252 Road of Batna, Khenchela 40004, Algeria}
	\affiliation{$^{3}$Center for Fundamental Physics, Zewail City of
		Science and Technology, Sheikh Zayed,12588, Giza, Egypt}
	\affiliation{$^{4}$School of Physics and Astronomy, University of
		Southampton, Highfield, Southampton SO17 1BJ, UK}
	\affiliation{$^{5}$Department of Physics and Astronomy, Uppsala University,
                     Box 516, SE-751 20 Uppsala, Sweden}
	\affiliation{$^{6}$Department of Physics, Bursa Uluda\~g University, TR16059 Bursa, Turkey }
          \affiliation{$^7$Departamento de Ciencias Integradas y Centro de Estudios Avanzados en F\'{i}sica Matem\'aticas y 
                    Computaci\'{o}n, Campus del Carmen, Universidad de Huelva, Huelva 21071, Spain}

\begin{abstract}
		
We investigate the ${\mathcal{R}}(D)$ and ${\mathcal{R}}(D^{\ast})$ anomalies in the context of the $B-L$ extension of the Minimal Supersymmetric Standard Model with Inverse Seesaw. We demonstrate that the lepton penguin $W^{\pm}l \bar\nu_l $ ($l=e,\mu,\tau$) mediated by CP-even/odd right-handed sneutrinos, charginos and neutralinos can account for these anomalies simultaneously.		

\end{abstract}
	
	\maketitle

\section{Introduction}
\label{sec:indtro}
In the many successes of the Standard Model (SM), the $B$-mesons and their decays play an important role. In particular, in addition to the observation of $B_{s}\rightarrow X_{s}\gamma$ and $B_{s}\rightarrow \mu^{+}\mu^{-}$ \cite{LHCb:2012skj,HFLAV:2022pwe} decays, the precise experimental measurements of their property provide an elegant way to determine the Cabibbo-Kobayashi-Maskawa (CKM) matrix elements \cite{LHCb:2020ist}. Furthermore, these decays are quite sensitive to  New Physics (NP) contributions, especially those happening through the transitions $b\rightarrow c l\bar{\nu}_{l}$ ($l=e,\mu,\tau$). Despite this dynamics occurs at tree-level in the SM, NP contributions at the same order of even at the loop level can be significant \cite{Fajfer:2012vx,Crivellin:2012ye,Crivellin:2013wna,Celis:2012dk,Celis:2016azn}. Also, compared with other semileptonic decays of $B$-meson, the $B\rightarrow D^{(*)} X$ ones are more advantageous since they are not CKM  suppressed and thus can be probed through many (differential) observables \cite{HFLAV:2019otj,Duraisamy:2014sna}. If Lepton Flavour Universality (LFU) is exact up to the lepton masses, the SM predicts the following Branching Ratios (BRs):  ${\rm BR}(B\rightarrow D\tau\bar{\nu_{\tau}}) \simeq 0.64\%$ and ${\rm BR}(B\rightarrow D^{*}\tau\bar{\nu_{\tau}}) \simeq 1.29\%$ \cite{Chen:2006nua}. Table \ref{tab:expRD} collects the results from the  LHCb, BaBar and Belle  collaborations reported between 2012 and 2020 in terms of the ratios
\begingroup
\small
\begin{equation}
\mathcal{R}(D)=\dfrac{{\rm BR}(B\rightarrow D\tau\bar{\nu_{\tau}})}{{\rm BR}(B\rightarrow Dl\bar{\nu_{l}})}, ~~ \mathcal{R}(D^{*})=\dfrac{{\rm BR}(B\rightarrow D^{*}\tau\bar{\nu_{\tau}})}{{\rm BR}(B\rightarrow D^*l\bar{\nu_{l}})}.
\label{eq:RDRDs}
\end{equation}
\endgroup
 Herein, the last row provides the combined results obtained by the Heavy Flavour Averaging  (HFLAV) group. 

\begin{table}
\hspace*{-0.75truecm}
{\setstretch{1.5}
\scalebox{0.9}{
\begin{tabular}{|c|l|l|}
\hline
 & \multicolumn{1}{c|}{$\mathcal{R}(D)$} & \multicolumn{1}{c|}{$\mathcal{R}(D^{*})$} \\ \hline 
SM & $0.299 \pm 0.003$ \cite{HFLAV:2019otj} & $0.258 \pm 0.005$ \cite{HFLAV:2019otj} \\ 
\hline
\multirow{2}{*}{LHCb} &  & $0.336 \pm 0.027 \pm 0.030$ \cite{LHCb:2015gmp}  \\ 
& & $0.283 \pm 0.019 \pm 0.029$ \cite{LHCb:2017smo, LHCb:2017rln} \\ \hline
\multirow{3}{*}{Belle} & $0.375\pm 0.064 \pm 0.026$ \cite{Belle:2019rba} & 
$0.283\pm 0.018 \pm 0.014$ \cite{Belle:2019rba} \\
& $0.307\pm 0.037 \pm 0.016$ \cite{Belle:2019rba} & $0.293\pm 0.038 \pm 0.015$ \cite{Belle:2015qfa} \\
& & $0.270\pm 0.035^{+0.028}_{-0.025}$ \cite{Belle:2017ilt,Belle:2016dyj} \\ \hline
BaBar & $0.440\pm 0.058 \pm 0.042$ \cite{BaBar:2012obs,BaBar:2013mob} & $0.332 \pm 0.024 \pm 0.018$ \cite{BaBar:2012obs,BaBar:2013mob} \\ \hline 
HFLAV & $0.339 \pm 0.026 \pm 0.014$ \cite{HFLAV:2022pwe} & $0.295 \pm 0.010 \pm 0.010$ \cite{HFLAV:2022pwe} \\ \hline
\end{tabular}}}
\caption{Experimental values for $\mathcal{R}(D)$ and $\mathcal{R}(D^{*})$ reported by the experimental  collaborations and HFLAV group.}
\label{tab:expRD}
\end{table}

The deviations between the experimental measurements and the SM predictions may hint at violation of LFU, which necessitates NP contributions \cite{Crivellin:2019dwb,Gomez:2019xfw,Marzo:2019ldg,He:2017bft,Iguro:2017ysu,Wei:2017ago,Celis:2012dk,Bhattacharya:2016mcc,Sakaki:2013bfa,Ko:2012sv,Crivellin:2012ye,Bhattacharya:2020lfm,Datta:2017aue,Bhattacharya:2014wla,Datta:2012qk,Fajfer:2012jt,Tanaka:2012nw,Alok:2017qsi,Alok:2019uqc}. One of the most promising  Beyond the SM (BSM) theories is Supersymmetry (SUSY). In the Minimal Supersymmetric SM (MSSM), its  minimal version, one may assume non-zero mixing among the slepton families to induce LFU violation at loop level. However, such a direct mixing is strictly constrained (see, e.g., \cite{Hammad:2016bng}) by Lepton Flavour Violation (LFV) experiments \cite{MEGII:2021fah,BaBar:2009hkt}. Nonetheless, the MSSM can still accommodate some deviations in $\mathcal{R}(D)$ and $\mathcal{R}(D^{*})$ through the penguin diagrams involving neutralinos, charginos and heavy Higgs bosons, but their contributions cannot fully recover the experimental measurements \cite{Boubaa:2016mgn,Boubaa:2020ksf,Hu:2020yvs}. In this paper, we surpass the MSSM  by assuming that SUSY is non-minimal \cite{Moretti:2019ulc}.

%
\begin{figure*}[!t]
\centering
\includegraphics[scale=0.45]{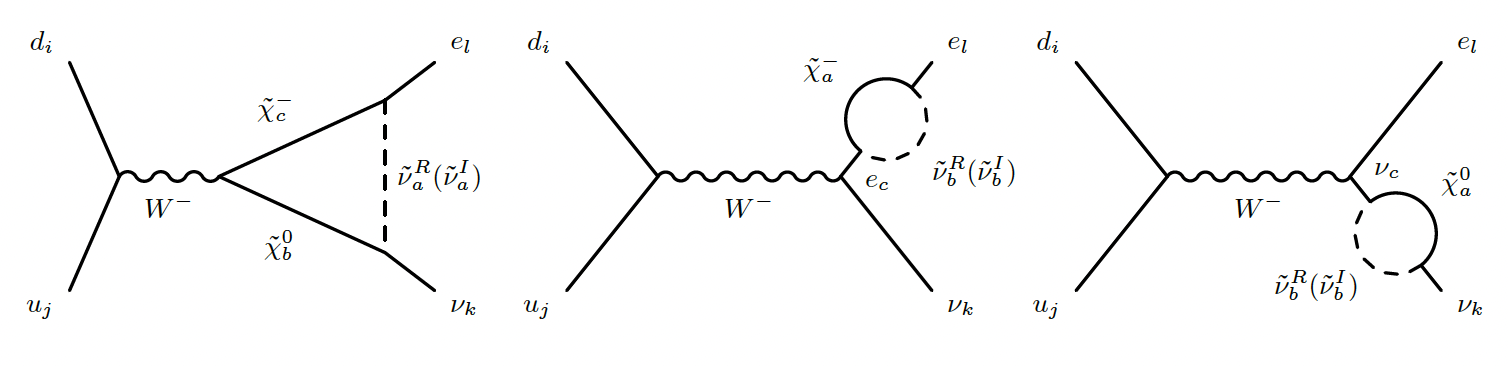}
\caption{Penguin and  self-energy diagrams in the BLSSM-IS contributing to the $b\rightarrow \bar{c}l\bar{\nu}_{l}$ transition.}
\label{fig:diags}
\end{figure*}
%

\section{The SUSY Model}
The $B-L$ extension of the MSSM with Inverse Seesaw (BLSSM-IS)
 is based on the gauge group $SU(3)_C \times SU(2)_L \times U(1)_Y \times U(1)_{B-L}, $ where 
$U(1)_{B-L}$ is spontaneously broken by two chiral singlet superfields 
$\hat{\chi}_{1,2}$ with $B-L$ charge $= \pm 1$. In addition to the MSSM superfields, a gauge boson $Z^{\prime}_{B-L}$ and  three chiral singlet superfields $\hat{\nu}^c_i$ with
$B-L$ charge $= -1$ are introduced for the consistency of the model. Finally, three  singlet fermions $S_1$ with $B-L$ charge $= -2$ and three singlet fermions $S_2$ with $B-L$ charge $= +2$ are employed to implement the IS mechanism \cite{Khalil:2010iu}. The superpotential of this model is given by%
\bea
W &=&  Y_u\hat{Q}\hat{H}_2\hat{U}^c + Y_d \hat{Q}\hat{H}_1\hat{D}^c+ Y_e\hat{L}\hat{H}_1\hat{E}^c \nonumber\\  &+&
Y_\nu\hat{L}\hat{H}_2\hat{\nu}^c+Y_S\hat{\nu}^c\hat{\chi}_1\hat{S}_2 +\mu\hat{H}_1\hat{H}_2+ \mu'\hat{\chi}_1 \hat{\chi}_2. 
\label{superpotential}
\eea
Electro-Weak Symmetry Breaking (EWSB) and $B-L$ radiative breaking at $v' = \sqrt{{v'_1}^2 +{v'_2}^2} \gsim 7 $ TeV are described in \cite{Khalil:2016lgy}. Here, we only consider the particle spectrum (masses and couplings)  relevant to our processes, i.e., the BLSSM-IS lightest  right-handed sneutrino and neutralino, with the lightest chargino being MSSM-like. If we write $\tilde{\nu}_{L}$, $\tilde{\nu}_{R}$ and  $\tilde{S}_2$ in terms of real and imaginary parts, one finds that  the mass of the lightest CP-odd sneutrino, $\tilde{\nu}^I_{1}$, is almost equal to that of the lightest CP-even one, $\tilde{\nu}^R_{1}$, and either   can be of  ${\cal O}(100$~GeV) \cite{Khalil:2015naa}.

The neutralinos $\tilde{\chi}^0_i $ ($i=1,\dots,7$)  in the BLSSM-IS are the physical (mass) superpositions of three fermionic partners of the 
neutral gauge bosons called gauginos $\tilde{B}$ (bino), 
$\tilde{W}^3$ (wino) and $\tilde{B'}$ ($B'$ino), in addition to the fermionic partners of both  the MSSM Higgs bosons ($\tilde{H}_1^0$ and $\tilde{H}_2^0 $) and  $B-L$ (pseudo)scalar bosons ($\tilde{\chi}_1$ and $\tilde{\chi}_2 $). The lightest neutralino has the following decomposition:
\be 
\tilde\chi^0_1\!=\!V_{11}{\tilde B}\!+\!V_{12}{\tilde
W}^3\!+\!V_{13}{\tilde H}^0_1\!+\!V_{14}{\tilde
H}^0_2\!+\!V_{15}{\tilde B'}\!+\!V_{16}{\tilde \chi_1}\!+\!V_{17}{\tilde
\chi_2}. 
\ee 
In addition to the typical MSSM gaugino or Higgsino, the Lightest Supersymmetric  Particle  (LSP) might be $B'$ino-like or $\tilde{\chi}_i$-like of order ${\cal O}(100$ GeV).
\section{Contributions to ${\cal R}(D)$ and ${\cal R}({D^*})$}
The effective Hamiltonian for $b \to c l\bar{\nu}_{l}$ is given by 
\begin{eqnarray}
\mathcal{H}_{eff} &=& \dfrac{4G_{F}V_{cb}}{\sqrt{2}}\left[ (1+g_{VL})[\bar{c}\gamma^{\mu}P_{L}b][\bar{l}\gamma_{\mu}P_{L}\nu_{l}] \right. \nonumber\\
&+& g_{VR}[\bar{c}\gamma^{\mu}P_{R}b][\bar{l}\gamma_{\mu}P_{L}\nu_{l}]+g_{SL}[\bar{c}P_{L}b][\bar{l}P_{L}\nu_{l}] \nonumber\\
&+& \left. g_{SR}[\bar{c}P_{R}b][\bar{l}P_{L}\nu_{l}] + g_{T}[\bar{c}\sigma^{\mu\nu}P_{L}b][\bar{l}\sigma_{\mu\nu}P_{L}\nu_{l}] \right],
\label{eq:Heff}
\end{eqnarray}
where $G_{F}$ is the Fermi constant, $V_{cb}$ is the CKM term which encodes the $b\rightarrow c$ transition, $g_{i}$ is a ratio of the Wilson coefficients defined as $g_{i}\equiv C_{i}^{{\rm SUSY}}/C_{i}^{{\rm SM}}$ with $i=VL,VR,SL,SR,T$ and $P_{L,R}$ are the projection operators. In our notation, $V,S,T$ stand for  vector, scalar and tensor while $L,R$~is the helicity state of the $b$-quark.

The obsevables $\mathcal{R}(D)$ and
$\mathcal{R}(D^*)$ can be defined as
\begin{equation}
\mathcal{R}(D)=\frac{\Gamma(\bar{B}\rightarrow D\tau\nu_{\tau})}{\Gamma(\bar{B}\rightarrow Dl\nu_{l})}=\frac{\int_{m_\tau^2}^{(m_B-m_D)^2}\frac{d\Gamma_{\tau}^D}{dq^2}dq^2}{\int_{m_l^2}^{(m_B-m_D)^2}\frac{d\Gamma^D_l}{dq^2}dq^2},
\end{equation}
\vspace*{-0.75truecm}
\begin{equation}
\mathcal{R}(D^{\ast})=\frac{\Gamma(\bar{B}\rightarrow D^{\ast}\tau\nu_{\tau})}{\Gamma(\bar{B}\rightarrow D^{\ast}l\nu_{l})}=\frac{\int_{m_\tau^2}^{(m_B-m_D^{\ast})^2}\frac{d\Gamma_{\tau}^{D^{\ast}}}{dq^2}dq^2}{\int_{m_l^2}^{(m_B-m_D^{\ast})^2}\frac{d\Gamma^{D^{\ast}}_l}{dq^2}dq^2}.
\end{equation}
The explicit dependence of ${\mathcal{R}}(D)$ and
${\mathcal{R}}(D^*)$ on the NP Wilson coefficients can be extracted by integrating the expressions for the differential decay rates in Refs.~\cite{Tanaka:2012nw,Sakaki:2013bfa}, where the helicity suppression effect  (squared light charged lepton mass ratio) is negligible, and  fix the form factors to their central values as in Ref.~\cite{HFLAV:2022pwe}. Therefore, one finds
\begin{align}
\mathcal{R}(D)&  =\frac{\Gamma_\tau^D}{\Gamma_e^D},~~~~\mathcal{R}(D^\ast)  =\frac{\Gamma_\tau^{D^\ast}}{\Gamma_e^{D^\ast}},
\end{align}
where
\begin{widetext}
	\begin{align}
	{\Gamma_\tau^D}&  =10^{-15}(2.632|g_{SR}^\tau+g_{SL}^\tau|^{2}%
	+2.810|1+g_{VL}^\tau +g_{VR}^\tau|^{2} +2.309 |g_{T}^\tau|^{2} +4 \operatorname{Re}[(1+g_{VL}^\tau+g_{VR}^\tau)(g_{SR}^\tau+g_{SL}^\tau)^{\ast
	}]\nonumber\\
	&+3.064\operatorname{Re}[(1+g_{VL}^\tau+g_{VR}^\tau)g_{T}^{\tau^{\ast}}]),\\
	{\Gamma_e^D}&  =10^{-15}(10|g_{SR}^e+g_{SL}^e|^{2}%
	+9.393|1+g_{VL}^e +g_{VR}^e|^{2} +6.293 |g_{T}^e|^{2} +6.755\times 10^{-3} \operatorname{Re}[(1+g_{VL}^e+g_{VR}^e)(g_{SR}^e+g_{SL}^e)^{\ast
	}]\nonumber\\
	&+8.559\times 10^{-3}\operatorname{Re}[(1+g_{VL}^e+g_{VR}^e)g_{T}^{e^{\ast}}]),\\
	{\Gamma_\tau^{D^\ast}}&  =10^{-14}(1.264\times 10^{-2} |g_{SR}^\tau%
	-g_{SL}^\tau|^{2}+0.511(|1+g_{VL}^\tau|^{2}+|g_{VR}^\tau|^{2})
	+8.570|g_{T}^\tau|^{2}+4.82\times 10^{-2}\operatorname{Re}[(1+g_{VL}^\tau+g_{VR}^\tau)\nonumber\\
	&\times(g_{SR}^\tau-g_{SL}^\tau)^{\ast
	}]+3.333\operatorname{Re}[g_{VR}^\tau g_{T}^{\tau\ast}]
	-2.278\operatorname{Re}[(1+g_{VL}^\tau)g_{T}^{\tau\ast}] -0.907\operatorname{Re}%
	[(1+g_{VL}^\tau)g_{VR}^{\tau\ast}),\\
	{\Gamma_e^{D^\ast}}&  =10^{-14}(7.566\times 10^{-2} |g_{SR}^e-g_{SL}^e|^{2}+2.033(|1+g_{VL}^e|^{2}+|g_{VR}^e|^{2})
	+34.807|g_{T}^e|^{2}+1.583\times 10^{-3}\operatorname{Re}[(1+g_{VL}^e+g_{VR}^e)\nonumber\\
	&\times(g_{SR}^e-g_{SL}^e)^{\ast
	}]+7.067\times 10^{-3}\operatorname{Re}[g_{VR}^eg_{T}^{e\ast}]
	-3.516\times 10^{-3}\operatorname{Re}[(1+g_{VL}^e)g_{T}^{\tau\ast}] -3.514\operatorname{Re}%
	[(1+g_{VL}^e)g_{VR}^{e\ast}]),
	\end{align}
\end{widetext}
where the above decay rates can be constrained by $\Gamma(W\rightarrow\tau\nu)/\Gamma(W\rightarrow l\nu)$ and $\Gamma(\tau \rightarrow \mu \nu_{\tau}\nu_{\mu})/\Gamma(\tau \rightarrow e \nu_{\tau}\nu_e)$. To determine the SM prediction for $\mathcal{R}(D)$ and $\mathcal{R}(D^\ast)$ the NP Wilson coefficients are set to zero, $g_i=0$.

The penguin corrections to the vertex $W^\pm l \nu_l $ $(l=e,\mu,\tau)$ yield the SUSY contributions to $g_{VL}$. These corrections are achieved in the MSSM by exchanging charginos, neutralinos and sleptons or left-handed sneutrinos. In the BLSSM-IS, the right-handed sneutrino with large $Y_\nu$ coupling can boost these contributions, as shown in Fig.~\ref{fig:diags}. The relevant Wilson coefficient is given by
\begin{eqnarray}        
\!\!C_{VL}^{\tilde\nu^{\rm R}} \!&= & \frac{\Gamma_{\tilde{\chi_b}^0 \nu_k \tilde\nu^R_a}^L \Gamma_{\bar{l_l}\tilde{\chi}_c^- \tilde\nu_a^R }^R \Gamma_{\bar{u_j}d_i W^-}^L}{16\pi^2m^2_{W^-}}  \big[-\Gamma_{\tilde{\chi}_c^-\tilde{\chi}_b^0 W^- }^L m_{\tilde{\chi}^0_{{b}}} m_{\tilde{\chi}^-_{{c}}} \nonumber\\
\!\!&\!-\!&\!C_0(m^2_{\tilde{\chi}^-_{{c}}}, m^2_{\tilde{\chi}^0_{{b}}}, m^2_{\tilde\nu_a^R}) \!+\! \Gamma_{\tilde{\chi}_c^-\tilde{\chi}_b^0 W^-}^R \big[B_0(m^2_{\tilde{\chi}^0_{{b}}}, m^2_{\tilde{\chi}^-_{{c}}}) \nonumber\\
\!\!&\!-\!&\! 2 C_{00}(m^2_{\tilde{\chi}^-_{{c}}}, m^2_{\tilde{\chi}^0_{{b}}}, m^2_{\tilde\nu_a^R}) \!+\! m^2_{\tilde\nu_a^R}C_0(m^2_{\tilde{\chi}^-_{{c}}}, m^2_{\tilde{\chi}^0_{{b}}}, m^2_{\tilde\nu_a^R})\!\big]\!\big]\!,~~~
\label{eq:WC}
\end{eqnarray}
with $R \to I$ for the CP-odd right-handed sneutrino. Here, $C_{0}(x,y,z)$, $B_{0}(x,y)$ and $C_{00}(x,y,z)$ are the loop functions defined  in \cite{Buras:2002vd} and $\Gamma$ stands for the coupling among the particles stated in the subindex. 

The $W$ decay to $l \nu$, whose ratio is subject to the  experimental constraints \cite{ParticleDataGroup:2016lqr}
\bea 
W_{\tau e} &\equiv& \dfrac{\Gamma(W\rightarrow\tau\nu)}{\Gamma(W\rightarrow e\nu)}  = 1.043 \pm 0.024, \nonumber\\
W_{\tau \mu} &\equiv& \dfrac{\Gamma(W\rightarrow\tau\nu)}{\Gamma(W\rightarrow\mu\nu)}  = 1.07 \pm 0.026,
\label{eq:Wleps}
\eea
may be affected by the mentioned penguin corrections. 

Along with the leptonic decays of $W$ boson, the precise measurements on the $\tau$ decays also play  an important role to control violations to  LFU. The following constraints are taken into account from experimental findings \cite{ParticleDataGroup:2016lqr}:
\begin{equation}
\tau_{\mu e} \equiv \dfrac{{\Gamma}(\tau \rightarrow \mu \nu_{\tau}\nu_{\mu})}{{\Gamma}(\tau \rightarrow e \nu_{\tau}\nu_{e})} = 0.979\pm 0.004~.
\label{eq:taumue}
\end{equation}
Furthermore, we apply the mass bounds on the SUSY spectrum \cite{ParticleDataGroup:2014cgo,ATLAS:2012yve,CMS:2012qbp,ATLAS:2017mjy,ATLAS:2019npw}, the constraints from the rare $B_{s}\rightarrow X_{s}\gamma$ and $B_{s}\rightarrow \mu^{+}\mu^{-}$ decay modes \cite{HFLAV:2022pwe} as well as the current limits on the LFV processes as $l_{i} \rightarrow l_{j}\gamma$ \cite{MEGII:2021fah,BaBar:2009hkt}. 

\section{Results}
\begin{figure*}[ht!]
\centering
\subfigure{\includegraphics[scale=0.45]{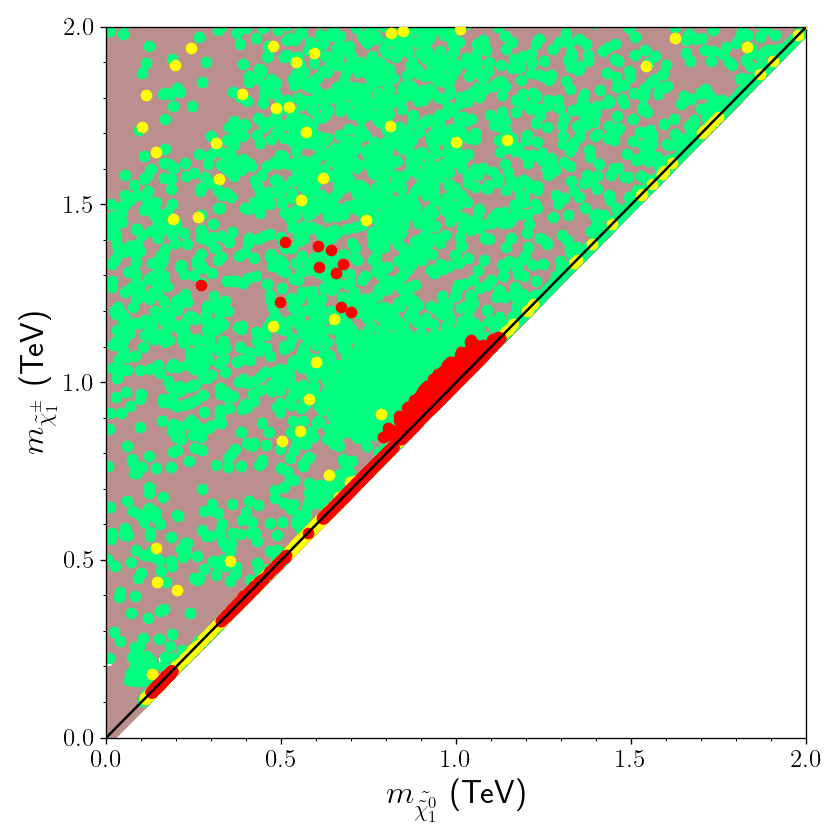}}%
\subfigure{\includegraphics[scale=0.45]{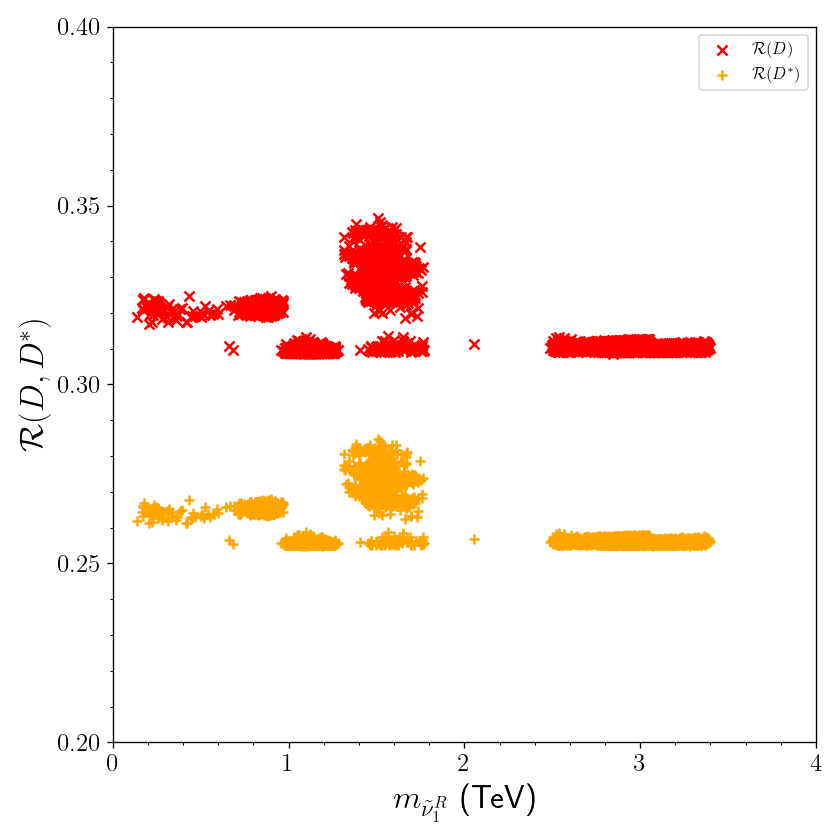}}
\caption{$\mathcal{R}(D)$ and $\mathcal{R}(D^{*})$  in correlation with the lightest chargino and neutralino (left) as well as right-handed sneutrino (right) masses. In the $m_{\tilde{\chi}_{1}^{\pm}}-m_{\tilde{\chi}_{1}^{0}}$ plane, all points are compatible with  EWSB. Green points also satisfy the SUSY mass bounds and the constraints on $B_{s}\rightarrow \mu^{+}\mu^{-}$ and $B_{s}\rightarrow X_{s}\gamma$. Yellow points indicate the solutions for $\mathcal{R}(D)$ and $\mathcal{R}(D^{*})$ within $2\sigma$  of the experimental measurements. The red points form a subset of the yellow ones as they  are also consistent with LFV and LFU constraints. The correlation with the right-handed sneutrino mass (common to CP-even and -odd states) is shown only for these red points.}
\label{fig:charneutsneut}
\end{figure*}

In this section, we display our results for $\mathcal{R}(D)$ and $\mathcal{R}(D^{*})$ consistent with the experimental constraints mentioned above. We used the Metropolis-Hastings algorithm, as described in  \cite{Baer:2008jn,Belanger:2009ti}, combined with SPheno  \cite{Porod:2011nf}, in turn generated with SARAH \cite{Staub:2013tta}, to scan the parameter space of the low scale BLSSM-IS. As some of the main contributions to $\mathcal{R}(D)$ and $\mathcal{R}(D^{*})$ come through the first diagram in Fig. \ref{fig:diags}, involving the lightest chargino and neutralino, one would naively expect to observe the largest corrections for
light masses of both of the latter. However, as shown in the left panel of Fig. \ref{fig:charneutsneut}, the experimental measurements of $\mathcal{R}(D)$ and $\mathcal{R}(D^{*})$ can be accommodated within $2\sigma$  even when these particles weigh around 2 TeV (yellow points). Alas, the LFV and LFU constraints exclude these solutions and allow only those (red points) with $m_{\tilde{\chi}_{1}^{\pm}}, m_{\tilde{\chi}_{1}^{0}} \lesssim 1$ TeV.

The reason why even relatively heavy masses for the  lightest chargino and neutralino can still accommodate the experimental measurements of $\mathcal{R}(D)$ and $\mathcal{R}(D^{*})$ is due to the fact that one of the main drivers of the NP corrections is the mass degeneracy between these two states. As seen from the red points over the $m_{\tilde{\chi}_{1}^{\pm}}-m_{\tilde{\chi}_{1}^{0}}$ plane in the same plot, the enhancement in $\mathcal{R}(D)$ and $\mathcal{R}(D^{*})$ mostly requires $m_{\tilde{\chi}_{1}^{\pm}} - m_{\tilde{\chi}_{1}^{0}} \simeq 0$. This degeneracy can be understood through the asymptotic behaviour of the vector-like Wilson coefficient. When two of the particles in the triangle loops are nearly degenerate in mass, the NP contributions become proportional to the mass of the third particle in it, up to some scales at which the (phase space)  suppression from the mass takes over \cite{Boubaa:2016mgn,Buras:2002vd}. Another fact leading to the mass degeneracy being instrumental to boost NP corrections is that the coupling $W-\tilde{\chi}^{\pm}-\tilde{\chi}^{0}$ takes its largest value when the chargino and neutralino are Wino-like and, indeed, the allowed SUSY spectra typically involve nearly mass degenerate Wino-like such states.

\begin{figure}
	\centering
	\includegraphics[scale=0.45]{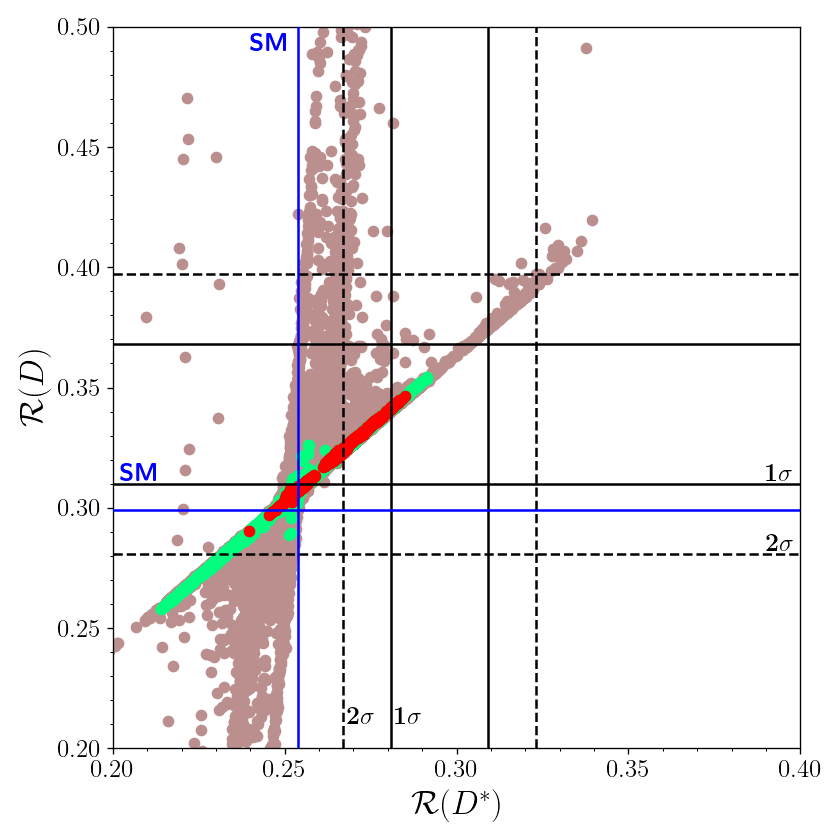}
	\caption{The correlation between $\mathcal{R}(D)$ and $\mathcal{R}(D^{*})$. Here, all points are compatible with  EWSB. Green points also satisfy the SUSY mass bounds and the constraints from $B_{s}\rightarrow \mu^{+}\mu^{-}$ and $B_{s}\rightarrow X_{s}\gamma$. The red points form a subset of the green ones they are also consistent with LFV and LFU constraints. The horizontal (vertical) black solid and dashed lines show the $1\sigma$ and $2\sigma$ ranges of the experimental measurements of $\mathcal{R}(D)$ ($\mathcal{R}(D^{*})$), respectively, while the blue line indicates the SM value.}
	\label{fig:RDRDs}
\end{figure} 

Considering the mass degeneracy and the asymptotic behaviour of the loop functions in the Wilson Coefficient given in Eq.~(\ref{eq:WC}), one can  then expect that also heavy lightest right-handed  sneutrino masses (essentially degenerate for  CP-even and -odd states) can yield a considerable enhancement in $\mathcal{R}(D)$ and $\mathcal{R}(D^{*})$. Indeed, the right  panel of Fig. \ref{fig:charneutsneut} shows that the $\mathcal{R}(D)$ and $\mathcal{R}(D^{*})$ results in our model are almost insensitive to the (common) right-handed sneutrino mass, so that compliance with experimental measurements within $2\sigma$ can be realised when this state  is as heavy as about 3.5 TeV.

Finally, we display the correlation between $\mathcal{R}(D)$ and $\mathcal{R}(D^{*})$ in Fig. \ref{fig:RDRDs}. Even though the theoretical solutions may not display any correlation (grey points), the experimental constraints allow only a linear relation between $\mathcal{R}(D)$ and $\mathcal{R}(D^{*})$, i.e., $\mathcal{R}(D) \simeq 1.2 \times \mathcal{R}(D^{*})$.

\section{Conclusions}
We have found that the BLSSM-IS is able to explain within $1\sigma$ the (averaged) measured values of $\mathcal{R}(D)$ and $\mathcal{R}(D^{*})$,
in presence of experimental constraints on its EW, SUSY and flavour sectors, notably including those from LFU and LFV observables. The additional NP contributions, above and beyond the SM ones, that enable this are given by penguin diagrams involving the lightest CP-even/odd right-handed sneutrino, neutralino and chargino. This result goes beyond what previously achieved for the MSSM, which is only partially able to comply with the observed values of these quantities. Specifically, it is noteworthy that ${\cal R}(D)$ is enhanced in the BLSSM-IS significantly more than in the MSSM. This is
due  to the following two reasons.
Firstly, the BLSSM-IS has additional right-handed sneutrinos (both  real and imaginary components) running in the relevant penguin diagram with,
owing to the IS dynamics for neutrino masses, large  Yukawa couplings between a charged Higgsino, charged lepton and such a  lightest right-handed 
sneutrino. Secondly, unlike in the MSSM, due to the significant contribution of such a right-handed sneutrino in the BLSSM-IS to both $\tau$ and light lepton ($\ell=e,\mu$) observables, $\Gamma_{\ell}$ can be suitably reduced with respect to its value in the SM. As overall result, the $\Gamma_{\tau}/\Gamma_{\ell}$ ratio that defines ${\cal R(D)}$ is thus improved. In summary then, $\mathcal{R}(D)$ (especially) and $\mathcal{R}(D^{*})$ (more moderately) appear to privilege a non-minimal realisation of SUSY.

\section*{Acknowledgments}
DB is supported by the Algerian Ministry of Higher Education and Scientific Research
under the PRFU Project No. B00L02UN400120230002. 
SK is partially supported by the Science, Technology and Innovation Funding Authority (STDF) under
Grant No. 37272. SM is supported in part through the
NExT Institute and the STFC Consolidated Grant No.
ST/L000296/1.
CSU is supported in part by the
Spanish MICINN, under grant PID2019-107844GB-C22.

\bibliographystyle{JHEP}
\bibliography{RDbib}

\end{document}